\documentclass[preprint,12pt]{elsarticle}
\usepackage{graphicx,color,xcolor}
\newcommand{\R}{{\mathbb{R}}}

\newcommand{\N}{{\mathbb{N}}}
\usepackage{amssymb}
\journal{Annals of Physics}

\begin{document}

\title{ Influence of a squeezed  prewell on tunneling 
properties and bound states in heterostructures 
 }
\author{Yaroslav Zolotaryuk and Alexander V. Zolotaryuk}
\address
{Bogolyubov Institute for Theoretical Physics, National Academy of
Sciences of Ukraine, Kyiv 03143, Ukraine}

\date{\today}

\begin{abstract}

A resonant tunneling effect of an extremely thin
potential well on the transmission of charged particles
 through a planar heterostructure with an arbitrary potential profile 
is investigated in a squeezing limit as the well width tends to zero.
In this limit, the transmission probability through the structure 
is shown to be blocked for all the parameter values of the well, except 
a resonance set of Lebesgue measure zero. 
The peak-to-valley ratio is shown to increase crucially with the 
squeezing of the well: the thinner is its thickness,
the resonant peaks become  sharper and  localized at isolated points. 
Contrary,  a discrete spectrum of the heterostructure (if any) 
does exist both on the resonance set and beyond it, however, the squeezing 
scenario here turns out to be quite interesting and sophisticated.

\end{abstract}

\maketitle

\bigskip

\section{Introduction}

In the fast developing field of nanoscience, the question how to control 
the particle transport in nanoscale quantum devices is very important. 
As a result, there are numerous theoretical studies on tunneling effects 
in one dimension describing the behavior of quantum particles when going 
through a potential barrier or passing across a potential well. 
In many cases, it is useful to perform these studies in the
zero-thickness limit when the barrier or well width shrinks to zero.
It is common knowledge that in the  barrier case, this limit
is realized as Dirac's delta function $\delta(x)$. However, for the positive-energy 
solutions of a one-dimensional stationary Schr\"{o}dinger equation
\begin{equation}
-\psi''(x) +V(x)\psi(x)= E\psi(x),
\label{1}
\end{equation}
where $\psi(x)$ is the wave function, $E$ the energy of quantum particles, 
$V(x)$ the potential for particles, and the prime stands for 
the differentiation over the space $x$, the similar limit of a  well-shaped potential
 $V(x)$ seems to have no relevance with a real
structure. In the present work, we argue that a more singular one-point 
limit, instead of the $\delta(x)$ distribution, has to be applied
in order to describe correctly the transmission properties of a quantum 
particle over a well. In this case, the squeezing rate of the well will exceeds
that of realizing a $\delta$-barrier and therefore one cannot expect any
well-defined limit function for the potential $V(x)$ in Eq.\,(\ref{1}). 
However, this is not a necessary condition, 
because the whole Schr\"{o}dinger equation may admit in the squeezing limit
a finite solution for the wave function $\psi(x)$. In this case, we are dealing
with a  so-called point interaction (see book \cite{a-h} for details 
as well as some recent works 
\cite{gnn,Lange0,Lange,m-cg,knt1,knt2,fggn}, a few to mention)
that provides the corresponding two-sided boundary conditions on the wave 
function $\psi(x)$ at a point of singularity \cite{adk}. 

The point limit of a quantum well materialized in the present paper appears to
be similar to that used for realizing the point interactions with 
potentials in the form of the derivative of Dirac's delta function
$\delta'(x)$. As established  
in the series of publications \cite{c-g,zci,tn,gm,gh1,gh2,g,zz14},
the main feature of the point interactions based on the  potential   $\gamma\delta'(x)$,
where $\gamma$ is a strength constant,  is the existence 
of a resonance set of isolated points in the $\gamma$-space, at which 
the transmission of quantum particles through a $\gamma\delta'$-barrier is non-zero,
whereas beyond the resonance points, this barrier acts as a perfectly reflecting wall. 
A similar point limit has recently been implemented for one-dimensional pseudospin-one
systems \cite{zzg}. 

An interesting problem concerns the perturbation of regular background 
potentials by different point interactions. One of these interactions is
a so-called $\delta'$-interaction for which the derivative 
 $\psi'(x)$ is continuous at the point of singularity, but $\psi(x)$ 
  discontinuous (the notation adopted in book~\cite{a-h}).
The self-adjoint Hamiltonian of a harmonic oscillator perturbed by this 
interaction has rigorously been defined in Ref.\,\cite{Albeverio1}.
  In a subsequent publication \cite{Albeverio2}, this study 
 has been extended for the perturbation by a
triple of $\delta'$-interactions using the Cheon-Shigehara approach 
\cite{Cheon}. Another type of the point interactions, which were used as a perturbation of background potentials, is materialized through the derivative 
of the delta-function $\delta'(x)$ (referred in 
the literature to as a $\delta'$-potential, see e.g., Ref.\,\cite{Nizhnik}).
This point interaction has been used as the perturbation of a constant electric
field and a harmonic oscillator \cite{Gadella1} as well as 
an infinite square well \cite{Gadella2}. The energy spectrum of a one-dimensional 
V-shaped quantum well perturbed by three types of point interactions 
composed from the $\delta$- and $\delta'$-potentials has been
studied  in the papers \cite{Fassari1,Fassari2,Fassari3}. 
All these systems are combined from 
a specific regular subsystem plus a point interaction centered inside 
this subsystem. It is of interest to examine the situation when a point 
interaction is found outside a background subsystem at some finite distance.

The present work focuses on the investigation of the role of 
a quantum well (referred in some publications to as a {\it prewell},
see, e.g., Refs.\cite{Boykin,Lewis,Pfenning}) 
on the tunneling of quantum particles through a given background structure 
being of a general form. The well is assumed to be extremely thin and located
apart from the background subsystem at some distance. 
 For our studies, it is convenient to use the transfer matrix approach. 
Some preliminaries concerning the calculation of scattering coefficients and 
 bound states,  using the transfer matrix for Eq.\,(\ref{1}) with
 an arbitrary potential $V(x)$, are presented in Section 2.
 In the next Section,  a point approximation of the well-shaped potential $V(x)$ 
in the limit as its support shrinks to zero, is examined. The comparison 
of this approximation with the $\delta$-limit is discussed.  
The transmission matrix of the bilayer composed of a rectangular well and
a layer with an arbitrary potential profile (referred to as a $B$-layer) 
is calculated in Section 4.
In the next Section, the influence of this squeezed well on the tunneling 
through the $B$-barrier is studied. Effects of the well on the discrete spectrum
of the $B$-layer are investigated  in Section 6. The next Section is devoted to
some concluding remarks.

\section{Scattering coefficients and bound states defined through transmission 
matrix elements }

Consider a potential profile $V(x)$  of an arbitrary shape
with the support on the interval $x_1 < x < x_2$, where $x_1$ and $x_2$ 
are arbitrary points on the $x$-axis. 
The transmission matrix for Eq.\,(\ref{1}) that connects the boundary conditions 
of the wave function $\psi(x)$ and its derivative $\psi^\prime(x)$
at $x=x_1$ and $x=x_2$ is defined through the matrix equation
\begin{equation}
\left(\begin{array}{ll} \psi(x_2) \\ \psi'(x_2) \end{array}
 \right) =\Lambda  
\left(\begin{array}{ll} \psi(x_1) \\ \psi'(x_1) \end{array}\right), \quad
\Lambda = \left( \begin{array}{ll} \lambda_{11}~~~\lambda_{12} \\
\lambda_{21}~~~\lambda_{22} \end{array} \right), 
\label{2}
\end{equation}
where $\lambda_{ij} \in \R$ and $\det\Lambda =1$.

For positive energy solutions of Eq.\,(\ref{1}) and the left-to-right current, 
the reflection-transmission coefficients $R$ and $T$ are defined using
the following scattering setting:
\begin{equation}
\psi(k;x)= \left\{ \begin{array}{ll} {\rm e}^{{\rm i}kx} +R{\rm e}^{-{\rm i}kx} 
& \mbox{for}~-\infty < x < x_1\,, \\
T {\rm e}^{{\rm i}kx}  & \mbox{for}~~x_2  < x < \infty, \end{array}\right.
\quad k :=\sqrt{E}\,,~~E>0.
\label{3}
\end{equation}
Using then these expressions as the boundary conditions at $x=x_1$ and 
$x =x_2$ in the matrix equation (\ref{2}), we obtain 
\begin{equation}
R = - \left[ \lambda_{11} - \lambda_{22} + {\rm i}\left(k \lambda_{12} 
+ k^{-1}\lambda_{21} \right)\right] D^{-1}{\rm e}^{2{\rm i}kx_1}, 
\quad T = 2 D^{-1} {\rm e}^{{\rm i}k(x_1-x_2)} ,  
\label{4}
\end{equation}
where $D := \lambda_{11} + \lambda_{22} -{\rm i} \left(k \lambda_{12} 
+ k^{-1}\lambda_{21}\right) $. Using that $\det\Lambda =1$, one can get the relation
 $|D|^2 = 4 +u^2 +v^2$, where 
\begin{equation}
 u := \lambda_{11} - \lambda_{22} \quad \mbox{and} \quad
v = k\lambda_{12} + k^{-1}\lambda_{21}\,.  
\label{5}
\end{equation}
Thus, the reflection and transmission probabilities can be expressed 
in terms of the matrix elements $\lambda_{ij}$ as follows
\begin{equation}
{\cal R }:= |R|^2 = {u^2 +v^2 \over 4 +u^2 +v^2 } \quad \mbox{and} \quad
{\cal T}:= |T|^2 = {4 \over 4 + u^2 +v^2}\,.
\label{6}
\end{equation}

Similarly, one can express the discrete energy spectrum 
in terms of the elements of the $\Lambda$-matrix (\ref{2}). To this end, 
consider the negative energy solutions of Eq.\,(\ref{1}) with  $E =- \kappa^2$,
where $\kappa >0$ describes the energy levels of the system. 
 In this case, beyond the interval $x_1 < x < x_2$ (a free-particle space),
the wave function is given by
\begin{equation}
\psi(\kappa ;x) = \left\{ \begin{array}{ll} C_1 {\rm e}^{\kappa (x-x_1)} & 
\mbox{for}~ -\infty <x < x_1\,, \\
C_2 {\rm e}^{-\kappa (x-x_2)} & \mbox{for}~~x_2 < x< \infty , \end{array}\right.
\label{7}
\end{equation}
where $C_1$ and $C_2$ are arbitrary constants. 
 Using the expressions (\ref{7}) as the boundary conditions at $x=x_1$ and 
$x =x_2$ in the matrix equation (\ref{2}), we obtain two equations 
with respect to the constants $C_1$ and $C_2$. Obeying the compatibility 
condition, we get a general  equation for bound states. It reads 
\begin{equation}
 \lambda_{11}  + \lambda_{22} + \kappa \lambda_{12} +  \kappa^{-1} \lambda_{21} =0, 
\label{8}
\end{equation}
where the matrix elements $\lambda_{ij}$'s in general depend on $\kappa$.

\section{A point approximation  of a potential well revisited} 
 
In general, a one-point approximation of an arbitrary potential $V(x)$ 
in Eq.\,(\ref{1}) can be realized by the replacement 
$V(x) \to V_\varepsilon(x) = \varepsilon^{-\nu}V(x/\varepsilon),$
 where   $\varepsilon >0$ is a dimensionless
squeezing parameter that tends to zero, and the dimensionless parameter $\nu >0$  
 has to be chosen appropriately in each particular case. In the typical case 
 as $\nu =1$, the  regular function $V_\varepsilon(x)$ converges to 
 $\alpha \delta(x)$ in the sense of distributions, 
 where $\delta(x)$ is Dirac's delta function 
 and the integral $\alpha = \int_{-\infty}^{\infty} V(x)dx$ becomes 
 a strength of the $\delta$-potential.
For values $\nu \neq 1$, the potential $V_\varepsilon(x)$ has no 
$\varepsilon \to 0$ limits in the sense of distributions. 
 
 Assume that the potential $V(x)$ in Eq.\,(\ref{1}) has the profile of 
a rectangular barrier or 
well, i.e. $V(x) \equiv V =$ const. $\neq 0$ on a finite interval 
$x_1 \le x \le x_2$ and $V(x) \equiv 0$
beyond this interval. For this potential, the $\Lambda$-matrix (\ref{2}) can be 
calculated explicitly, resulting in
\begin{equation}
\Lambda = \left(\!\! \begin{array}{ll} ~~~\cos(ql)~~~~~~~~q^{-1}\sin(ql) \\
- q \sin(ql)~~~~~~~~\cos(ql) \end{array} \right), \quad l:=x_2 -x_1\,, 
\quad q = \sqrt{E - V}.
\label{9}
\end{equation}
Then, the calculation of the transmission probability ${\cal T}$
according to the second formula (\ref{6}) leads to the expression
\begin{equation}
{\cal T} = \left[ 1+ {V^2 \over 4E(E-V)}\sin^2(ql)\right]^{\!-1} \! .
\label{10}
\end{equation} 
In the particular case of tunneling through a rectangular barrier with
height $h$ and width $l$, the delta-function approximation 
($\nu =1$) is specified as 
 $V = \varepsilon^{-1} h$ and $l=\varepsilon a$. Then, in the limit as 
$\varepsilon \to 0$,  the $\delta$-potential strength is $\alpha = ha$
and the formula (\ref{10}) reduces to the standard form
 \begin{equation}
{\cal T} =  {1 \over 1 +(\alpha/2k)^2}
\label{11}
\end{equation}
that describes a monotonic decrease of the tunneling transmission with the growth 
of the strength $\alpha$. 
 
\begin{figure}[htb]
\begin{center}
\includegraphics[width=0.75\columnwidth]{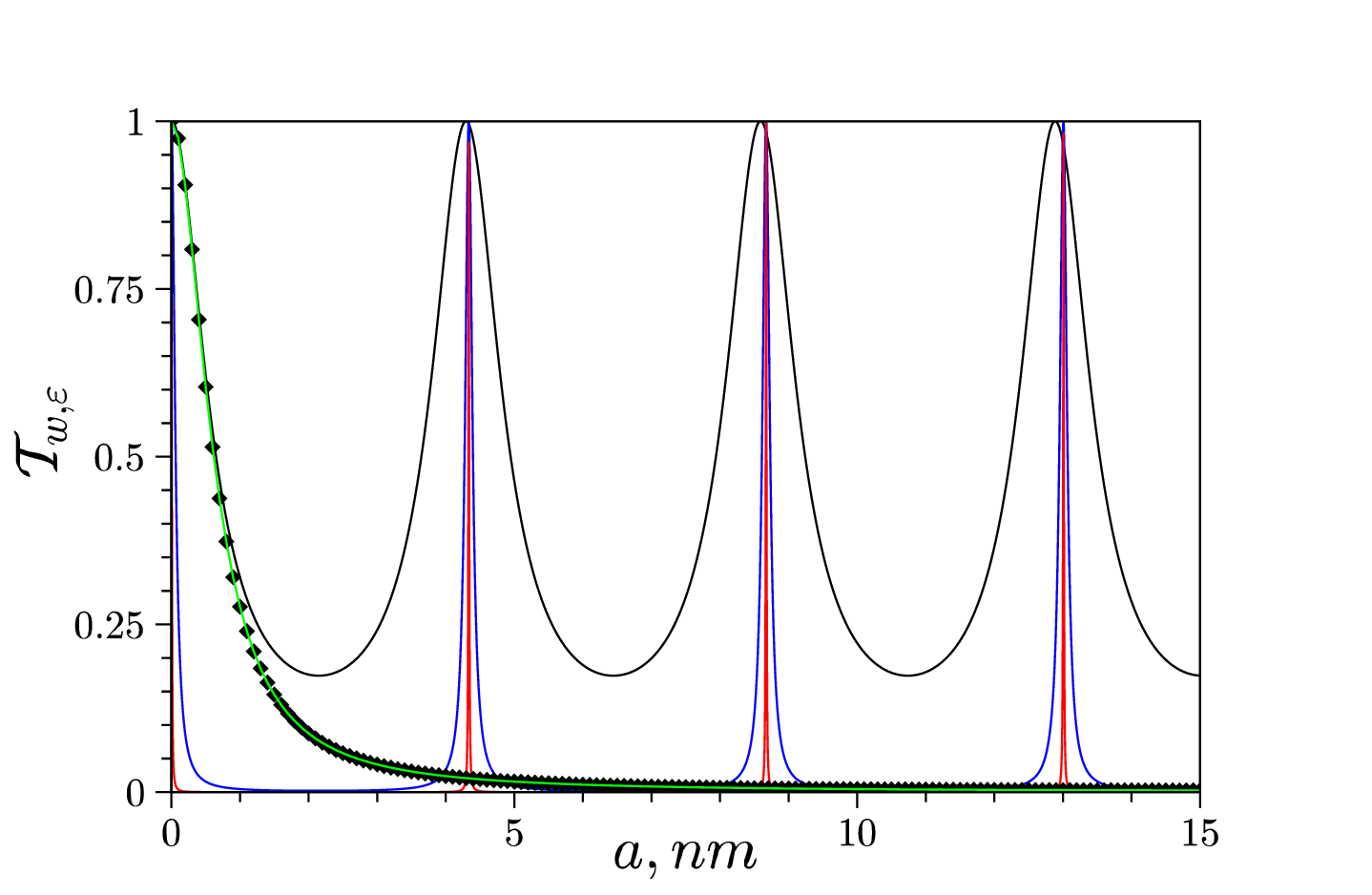}
\end{center}
\caption{\label{fig1} 
The transmission probability ${\cal T}_{w, \varepsilon}$ as a function of
the thickness parameter $a$ plotted according to formula (\ref{12a}) with 
 $E = 0.01$\,eV and $d =0.2$\,eV. The transmission
  probability for the unsqueezed (realistic) potential
($\varepsilon =1$) is shown by black line. The squeezing  
  ($\varepsilon \to 0$) scenario in the case $\nu =2$ is illustrated 
  by the lines plotted for   $ \varepsilon =0.1$ (blue) and  
  $ \varepsilon =0.01$ (red). The green line together with  $\blacklozenge$
 corresponds to the $\delta$-approximation ($\nu =1$) with $\varepsilon=0.01$,
 calculated according to formula (\ref{10}), where $E = 0.01$\,eV,
  $V=\varepsilon^{-1}h$, $h=0.2$\,eV,   and $l=\varepsilon a$.
 The calculations have been carried out
in the units for which $\hbar^2/2m^* =1$. Here, $m^*$  is an effective
electron mass chosen to equal $0.1\,m$, where $m$ is the free electron mass, 
so that 1\,eV = 2.62464\,nm$^{-2}$. 
}
\end{figure}

 However, in the case of the transport of particles across a potential 
 well, instead of 
 the monotonic law (\ref{11}), we are dealing with an oscillating behavior of the 
 transmission (\ref{10}) owing to the factor $\sin^2(ql)$ where $q$ is real.
Therefore the $\delta$-well approximation by the point potential
 $V(x)=-\alpha \delta(x)$, where 
$\alpha = ad$ ($d$ is a well depth), fails to describe adequately the transmission 
given by the same formula (\ref{11}). 
 Instead, one can use in the formula (\ref{10}) a more singular approximation
 (using $\nu =2$), 
 namely $ V = - \varepsilon^{-2}d$, replacing $q$ and $l$ defined in (\ref{9}) with 
\begin{equation}
q = q_\varepsilon = \sqrt{E + \varepsilon^{-2}d} \quad \mbox{and} \quad
l=\varepsilon a.
\label{12}
\end{equation}  
Within this approximation, the transmission probability
\begin{equation}
{\cal T}={\cal T}_{w,\varepsilon}=
\left[ 1 + {d^2 \over 4\varepsilon^2 E(\varepsilon^2 E +d)}
\sin^2\!\left( \sqrt{\varepsilon^2 E+d}\,a\right)\right]^{\!-1}
\label{12a}
\end{equation}
 exhibits an oscillating behavior for $\varepsilon >0$,
where the resonant peaks become more pointed as $\varepsilon \to 0$. 
The $\varepsilon \to 0$ limit of (\ref{12a}) reads
\begin{equation}
{\cal T} = {\cal T}_{w} = \lim_{\varepsilon \to 0} {\cal T}_{w,\varepsilon} 
 =\left\{ \begin{array}{ll} 1 & \mbox{if}~~ \sqrt{d}\,a 
 =n\pi, \\ 0 & \mbox{if}~~  \sqrt{d}\,a \neq n\pi,\end{array} \right. 
 \quad n=1,\,2,\, \ldots . 
\label{14}
\end{equation}
Thus, the transmission probability (\ref{14}) appears 
 to be non-zero only on some lines in the $\{d,a\}$-space
 forming a set of Lebesgue measure zero given by 
\begin{equation}  
  \Sigma := \{d,a~ \vert ~\sqrt{d}\,a=n\pi,~n \in \N \}
\label{15} 
\end{equation} 
and referred from now on a {\it resonance} set. Beyond this set,
 the transmission is identically zero. Therefore  the zero-range approximation 
 with the scaled function
\begin{equation} 
  V_\varepsilon(x) = -\left\{ \begin{array}{ll} \varepsilon^{-2} d &
  \mbox{for}~~0 \le x \le l=\varepsilon a, \\ 0 & \mbox{otherwise} \end{array}
  \right.
\label{16}
\end{equation}  
 turns out to be more adequate 
 than the $\delta$-approximation because it exhibits peaks and valleys 
 existing in the realistic case ($\varepsilon =1$), as shown in   
 Fig.\,\ref{fig1}. Contrary to the formula (\ref{11}), this figure 
 illustrates the  {\it spire-like} convergence of the transmission
  probability ${\cal T}_{w,\varepsilon}$ as $\varepsilon \to 0$.  

{\it Summary 1:~Let the Schr\"{o}dinger equation (\ref{1}),
in which the shape of the potential $V(x)$ is a rectangular well with depth $d$ 
and width $a$, is used for the description of the transmission of quantum particles  
across the well. For realizing a well-defined and adequate point interaction, 
the rectangular approximation (\ref{16}) 
 has to be applied  instead of the $\delta$-well approximation. For the 
  approximation (\ref{16}), in the squeezing limit as 
 $\varepsilon \to 0$,  the spire-like convergence of the transmission 
${\cal T}_{w,\varepsilon}$  leads to the existence of a non-zero (being perfect) 
transmission only on isolated lines forming the resonance set (\ref{15}). 
Beyond this set, the well acts as a perfectly reflecting wall.
}

\section{The special case of a $W$--$B$ bilayer structure}

\begin{figure}[htb]
\begin{center}
\includegraphics[width=0.75\columnwidth]{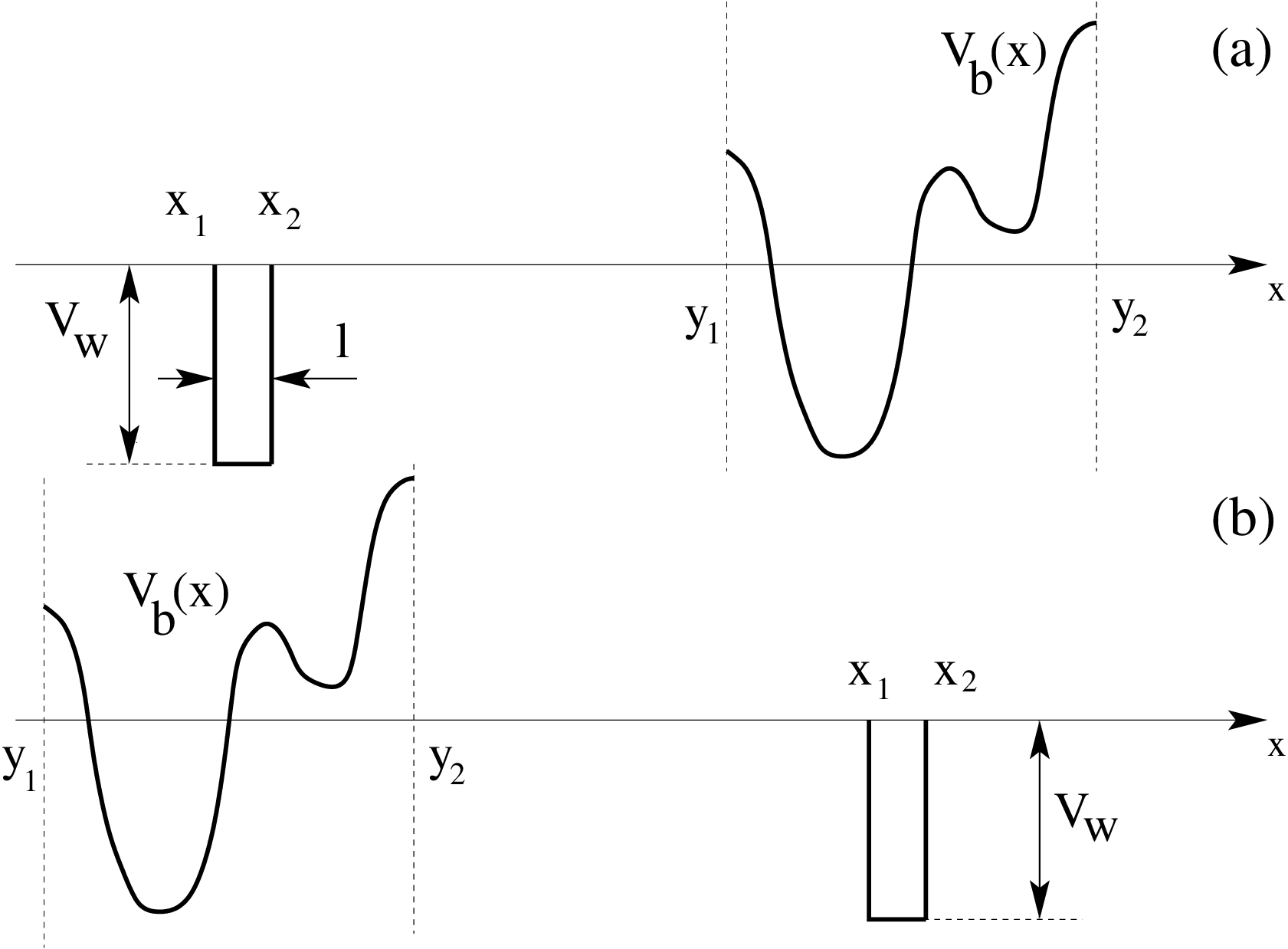}
\end{center}
\caption{\label{fig2} 
Sketch of the two cases of a $W$--$B$ bilayer system. 
(a) A rectangular well (emitter prewell) is located in front 
of the $B$-layer with an arbitrary potential profile. 
(b) A rectangular well (collector prewell) is found behind the $B$-layer.    
}
\end{figure}

In the following we will consider a bilayer heterostructure composed 
of a  layer with the potential in the form of a rectangular well 
(also called in some papers a  prewell, see, e.g., 
 Refs.\,\cite{Boykin,Lewis,Pfenning})
and a background layer with the profile of an arbitrary shape. Schematically,
this structure is shown in Fig.\,\ref{fig2} 
regarding the two situations:
the prewell (also referred to as a $W$-layer or a $W$-subsystem)
is incorporated either in front of the background layer 
(called a $B$-layer or a $B$-subsystem) or behind it.

Assume that the transmission matrix $\Lambda= \Lambda_w$
given by Eq.\,(\ref{9}) corresponds to the $W$-layer with 
the potential profile  $V_w(x) \equiv V <0$ 
defined  on the  interval $x_1 \le x \le x_2$ and 
 $V_w(x) \equiv 0$ beyond this interval. Suppose further that 
 the potential $V_b(x)$ for   the $B$-layer is defined on another interval 
 $y_1 \le x \le y_2$ and the elements of the corresponding transmission
matrix $\Lambda_b$ are denoted by $\lambda_{ij}$'s. Let the distance
$\rho$ between the $W$- and $B$-layers be non-zero. Then the transmission matrix
that corresponds to the free space between the subsystems is
\begin{equation}
\Lambda_0 = \left( \begin{array}{cc} ~~~\cos(k\rho)~~~~~k^{-1}\sin(k\rho) \\
- k \sin(k\rho)~~~~~~~\cos(k\rho) \end{array}~ \right), \quad k :=\sqrt{E}\,.
\label{17}
\end{equation}
The transmission matrix of the whole $W$--$A$ system is the product 
\begin{equation}
C = \left( \begin{array}{ll} c_{11}~~~c_{12} \\
c_{21}~~~c_{22} \end{array} \right)= \left\{ \begin{array}{ll} C_{wb} :=
\Lambda_b\Lambda_0\Lambda_w & \mbox{for}~x_1 < x_2 < y_1 < y_2\,, \\
C_{bw} := \Lambda_w\Lambda_0\Lambda_b & \mbox{for}~y_1 < y_2 < x_1 < x_2\,,
\end{array} \right.
\label{18}
\end{equation}
where the matrix $C_{wb}$ corresponds to the ($w-b$)-configuration 
 (the prewell is located in front of the $B$-potential and $y_1 - x_2 =\rho$)
   and the matrix $C_{bw}$ to the ($b-w$)-configuration (the prewell is found behind 
the $B$-potential and $x_1 - y_2 =\rho$). Explicitly, for the ($w-b$)-configuration, 
the $c_{ij}$-elements  are given by
\begin{equation}
\begin{array}{llll} \medskip
c_{11} = c_{11; \, wb} &=& \lambda_{11} \left[ \cos(q l) \cos(k\rho)
- (q /k)\sin(ql) \sin(k\rho) \right] \\ \medskip
&-& \lambda_{12}\left[q \sin(q l) \cos(k\rho) + k\cos(q l) 
\sin(k\rho) \right], \\ \medskip
c_{12} = c_{12;\, wb} &=& \lambda_{11}\left[q^{-1} \sin(ql) \cos(k\rho)
+ k^{-1}\cos(q l) \sin(k\rho) \right] \\ \medskip
&+& \lambda_{12}\left[\cos(q l) \cos(k\rho)- (k/q)\sin(ql) \sin(k\rho) \right],
\\ \medskip
c_{21} = c_{21; \, wb} &= &\lambda_{21}\left[ \cos(q l) \cos(k\rho)
- (q /k)\sin(q l) \sin(k\rho) \right] \\ \medskip
&-& \lambda_{22}\left[q\sin(ql) \cos(k\rho) + k\cos(q l) \sin(k\rho) \right], 
\\ \medskip
c_{22} = c_{22;\, wb} &=& \lambda_{21}\left[ q^{-1}\sin(q l) \cos(k\rho)
+ k^{-1}\cos(q l) \sin(k\rho) \right] \\ \medskip
&+& \lambda_{22}\left[\cos(ql) \cos(k\rho) - (k/q)\sin(ql) \sin(k\rho) \right].
\end{array}
\label{19}
\end{equation}
For the ($b-w$)-configuration, the matrix $C_{bw}$ is obtained from the 
matrix elements (\ref{19}) by the replacements 
$\lambda_{11} \longleftrightarrow \lambda_{22}$ as follows
\begin{equation} 
C_{bw}= \left( \begin{array}{lr} \medskip
c_{22;\, wb}(\lambda_{22} \to \lambda_{11})~~~
c_{12; \,wb}(\lambda_{11} \to \lambda_{22}) \\
c_{21;\, wb}(\lambda_{22} \to \lambda_{11}) ~~~
c_{11; \,wb}(\lambda_{11} \to \lambda_{22})\end{array} \right).
\label{20}
\end{equation}
Below we will study the influence of a squeezed prewell on a 
$B$-subsystem in two situations. First, it is of interest to know how
the presence of a prewell will act on  the transmission
of quantum particles through a potential barrier.
Second, the influence of a prewell on the structure of 
bound states in a $B$-well is to be investigated.
To implement the squeezing limit,  in  Eqs.\,(\ref{19}) and (\ref{20}),
we use  the $\varepsilon$-dependence (\ref{12}).  In this limit, the distance 
$\rho$ and the width $l_b :=y_2-y_1$ are accepted to be fixed.

\section{Influence of a squeezed prewell on the tunneling through 
an arbitrary barrier}

Assume in this section that the potential $V_b(x)$ has a non-zero barrier part
and a rectangle-like prewell is incorporated at distance $\rho$  
either in front of the $B$-barrier or behind it.
Intuitively, due to the resonant transmission (\ref{14}), 
one can expect that the squeezed prewell will allow a net tunneling
current through the barrier only on the resonance set $\Sigma$. 
More rigorously, based on Eqs.\,(\ref{19}) and (\ref{20}),  this can be proven
as follows.  Setting in these equations $q $ and $l$ given by 
(\ref{12}), we find that the elements $c_{11}$ and $c_{21}$ 
in general diverge as $\varepsilon \to 0$. However, due to the factor 
$\sin(ql)$, this divergence is suppressed on the set $\Sigma$.
More precisely, in the limit as $\varepsilon \to 0$, we have 
\begin{equation}
\! q|\sin(ql)| =\varepsilon^{-1} \sqrt{\varepsilon^2 E+d}\,\Big | \sin\!
\left(\!\sqrt{\varepsilon^2 E+d}\,a \right) \!\Big |\to  \left\{ \begin{array}{ll}
\, 0 & \mbox{for} ~(d,a) \in \Sigma, \\ \infty & \mbox{for}~(d,a) \notin
\Sigma . \end{array} \right. \!\!
\label{20a}
\end{equation}
As a result, for
the expressions (\ref{5}), where $u=u_\varepsilon$ and $v=v_\varepsilon$ 
are used for the whole system, i.e.,  
$\lambda_{ij}$'s are replaced respectively by $c_{ij}$'s given by 
Eqs.\,(\ref{19}) and (\ref{20}), we obtain the $\varepsilon \to 0$ limits  
\begin{equation}
\left\{ \begin{array}{ll} \smallskip u_\varepsilon \\ 
v_\varepsilon \end{array} \right\} \to (-1)^n \!\left\{ \! \begin{array}{ll} 
\smallskip \left[  (\lambda_{11} - \lambda_{22})\cos(k\rho) \mp \left(k\lambda_{12}
+ k^{-1}\lambda_{21}\right ) \sin(k\rho)\right] \\ 
 \left[ (k\lambda_{12} + k^{-1}\lambda_{21} ) \cos(k\rho)
 \pm  (\lambda_{11} - \lambda_{22} ) \sin(k\rho)\right]  \end{array} \!\right\}
\label{21}
\end{equation}
on the set $\Sigma$, whereas $|u_\varepsilon|$ and $|v_\varepsilon| \to \infty$
outside the set $\Sigma$. Here, 
 the upper signs in the square brackets belong to the $(w-b)$-configuration
and the lower ones to the $(b-w)$-configuration. 
From both these equations we immediately obtain the same expression for 
$u^2 +v^2$ if the prewell would be absent. Thus, the total tunneling transmission
through the whole system coincides with the transmission
across the barrier described by the $\Lambda_b$-matrix, 
however, this occurs only on the resonance set $\Sigma$ for the squeezed prewell.  

{\it Summary 2:  Let ${\cal T}_{wb}$ or $ {\cal T}_{bw}$ be the transmission 
probability of quantum particles tunneling through the system composed of 
a squeezed (as $\varepsilon \to 0$) prewell located
at any distance $\rho$ from a barrier (in front of the barrier or behind it). 
 This transmission is  $ {\cal T}_{wb}= {\cal T}_{bw} =  {\cal T}_{w} \cdot {\cal T}_{b} $
where  ${\cal T}_{w}$ is defined by (\ref{14}).  Moreover, it does not depend on the 
distance $\rho$.
}
 
\begin{figure}[htb]
 \begin{center}
\includegraphics[width=0.49\textwidth]{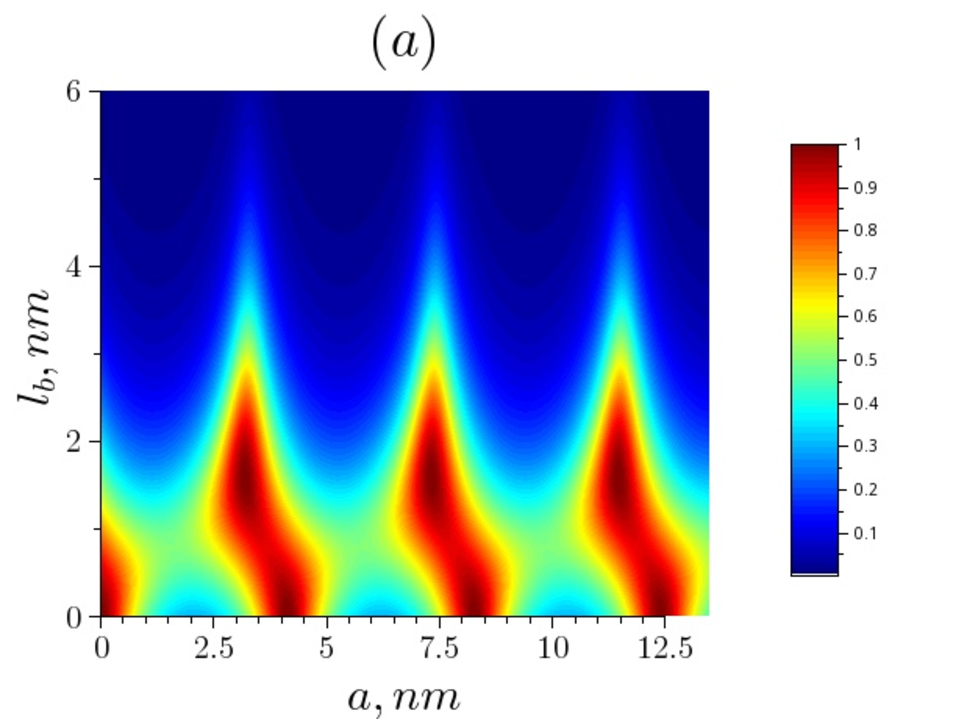}
\includegraphics[width=0.49\textwidth]{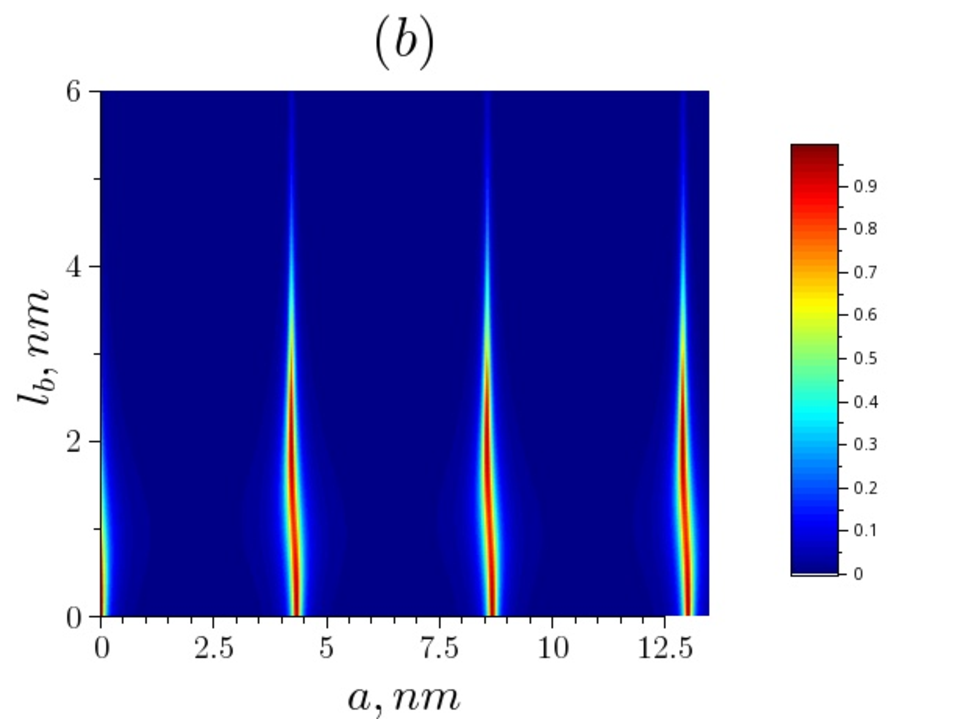}
\end{center}
\caption{\label{fig3}
The transmission probability ${\cal T}_{bw,\varepsilon}$ as a function of widths $a$ and $l_b$
for the  system composed of prewell with depth $d=0.2$\,eV and barrier with height
$V_b =0.1$\,eV, which has been calculated for two
values: (a) $\varepsilon = 1$ and (b) $\varepsilon = 0.1$.
Here, $E=0.02$\,eV and $\rho =10$\,nm. Panel (a) 
corresponds to $\varepsilon =1$, while panel (b) to $\varepsilon =0.1$.
The calculations have been performed using the same units as in 
Fig.\,\ref{fig1}. }
\end{figure}

This conclusion can be illustrated on the  simple example of a barrier subsystem. 
The transmission probability for this subsystem is a monotonically 
decreasing function given by the formula (\ref{11}).
The presence of an additional well crucially changes this transmission making it
resonant and dependent on the well parameters as demonstrated by the contour 
plot for the ($b-w$)-configuration in Fig.\,\ref{fig3}.
 Here the widths $a$ and $l_b$ are variables, and distance $\rho$ is arbitrary and 
fixed. The plots are present  for the whole system at two values of
$\varepsilon$ shown in Fig.\,\ref{fig3}(a) for $\varepsilon =1$ 
(unsqueezed prewell) and in Fig.\,\ref{fig3}(b) for 
$\varepsilon = 0.1$ (squeezed prewell). The figure clearly demonstrates
that the presence of a prewell essentially improves the resonance properties 
of the transmission through a barrier.

\section{Effects of a squeezed prewell on the discrete spectrum of 
bound states in the $B$-layer of an arbitrary profile}

In this section, we consider the situation  when Eq.\,(\ref{1}) 
with a potential $V_b(x)$ (the prewell is absent) admits
 the existence of negative-energy solutions with  
$E =- \kappa^2$, where $\kappa >0$ describes the discrete spectrum of 
bound states. For this case, we assume that the equation for the bound state 
levels $\kappa$  is given by Eq.\,(\ref{8}), where 
the elements $\lambda_{ij}$'s correspond to the 
potential $V_b(x)$ and consequently to the $\Lambda_b$-matrix. 
Similarly, for the bound states of the whole $W$--$A$ system  defined on the interval 
$(0,\, l+\rho +l_b)$, we have the same Eq.\,(\ref{8}), in which $\lambda_{ij}$'s
are replaced with $c_{ij}$'s given by Eqs.\,(\ref{19}) and (\ref{20})
where $k={\rm i}\kappa$,
$q$ and $l$ are determined by Eqs.\,(\ref{12}) with $E=-\kappa^2$, 
i.e., $q = q_\varepsilon = \sqrt{\varepsilon^{-2}d- \kappa^2}$ and $l=\varepsilon a$.
  Explicitly, for the ($w-b$)-configuration, Eq.\,(\ref{8}) with respect to
  the unknown $\kappa$ becomes
\begin{eqnarray}
&& \!\!\! \left(\lambda_{11} + \lambda_{22} + \kappa  \lambda_{12} 
+ { \lambda_{21} \over \kappa } \right) (1 +\tau) 
+ \left( {\kappa \over q_\varepsilon} \lambda_{11} 
- {q_\varepsilon \over \kappa } \lambda_{22} 
 -q_\varepsilon \lambda_{12}  
 + {\lambda_{21} \over q_\varepsilon} \right) t_\varepsilon  \nonumber \\
 &+&\!\!\! \left ({ \kappa^2 \over  q_\varepsilon}\lambda_{12} 
 -{q_\varepsilon \over  \kappa^{2}} \lambda_{21} 
 -{ q_\varepsilon \over \kappa } \lambda_{11} 
 + {\kappa \over q_\varepsilon} \lambda_{22} \right)  t_\varepsilon \tau=0,
 \label{22}
\end{eqnarray}
where  $t_\varepsilon := \tan(q_\varepsilon l)= 
\tan\!\left(\sqrt{d- (\varepsilon \kappa)^2}\,a\right)$ and $\tau := \tanh(\kappa \rho)$.
The equation for the ($b-w$)-configuration is obtained from 
Eq.\,(\ref{22}) by the replacements $\lambda_{11}  
 \longleftrightarrow  \lambda_{22}$\,. On the resonance set $\Sigma$,
 we have $t_\varepsilon \to 0$ and therefore Eq.\,(\ref{22}) 
 reduces to Eq.\,(\ref{8}) 
  that describes the discrete spectrum of the unperturbed $B$-subsystem. 
Beyond the set $\Sigma$, we have 
$\lim_{\varepsilon \to 0}t_\varepsilon \neq 0$, however, in this case,
 $q_\varepsilon \to \infty$ and therefore Eq.\,(\ref{22}) asymptotically reduces to
\begin{equation}
\lambda_{11}\tau +\lambda_{22} + \kappa \lambda_{12} 
+\kappa^{-1}\lambda_{21}\tau =0
\label{23}
\end{equation}
for the ($w-b$)-configuration. Replacing here $\lambda_{11}  
 \longleftrightarrow  \lambda_{22}$\,, one obtains  the equation
for the ($b-w$)-configuration. Therefore, in the $\varepsilon \to 0$ limit,
outside the resonance set $\Sigma$,  the bound state energies depend only 
on the distance $\rho$ through  the factor $\tau$. For sufficiently large
distances $\rho$, we have $\tau \to 1$ and therefore Eq.\,(\ref{23}) 
 coincides with Eq.\,(\ref{8}). This means that the squeezed 
prewell does not affect the discrete spectrum of the $B$-subsystem
if the $W$- and $B$-subsystems are  separated by sufficient large $\rho$.
Note that the squeezed $W$-well itself has no bound states. 

\begin{figure}[htb]
\begin{center}
\includegraphics[width=0.74\columnwidth]{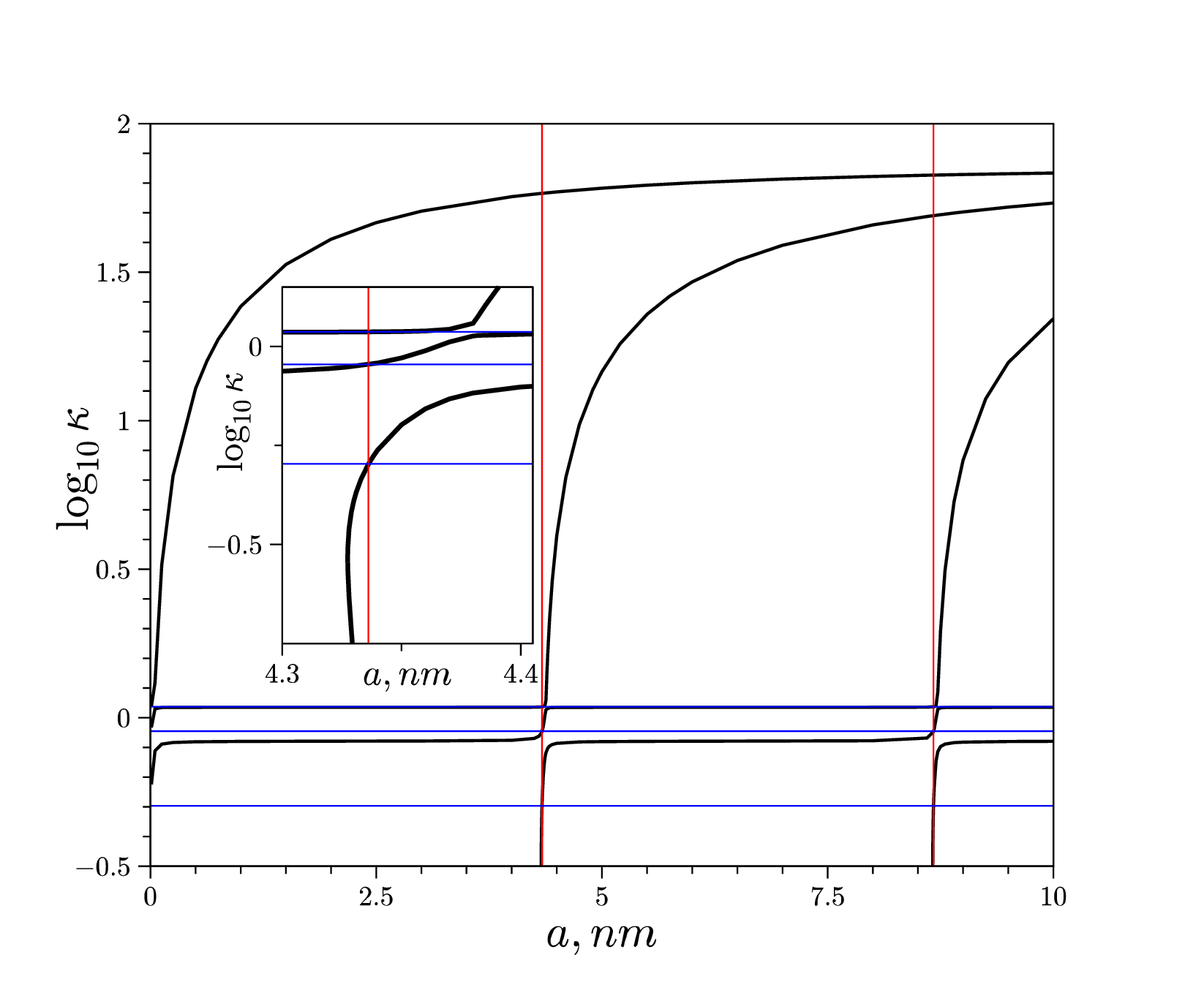}
\end{center}
\caption{\label{fig4} 
Dependence of the discrete spectrum in the case $N= 3$ on  the
 thickness parameter $a$ of the $W$-well 
 for $\varepsilon =0.01$, which is shown by solid (black) curves. 
The parameter values are  $d =0.2$\,eV, $\rho=0.5$\,nm,  $V_b= -\, 0.5$\,eV and $l_b=7$\,nm.  
Solutions to Eq.\,(\ref{8}): $  \kappa_1^0 = 1.08819, \, \kappa_2^0= 0.90138,\, 
\kappa_3^0= 0.50528 $ are  indicated  by blue horizontal straight lines. 
Between the origin $a=0$ and 
the first resonance point $a=a_1$, the spectrum consists of three levels 
$\kappa_{1,\varepsilon}$\,, 
$\kappa_{2,\varepsilon}$ and $\kappa_{3,\varepsilon}$ ($\kappa_{1,\varepsilon}> 
\kappa_{2,\varepsilon} >\kappa_{3,\varepsilon}$). At  point $a=a_1$, 
a new level $\kappa_{4,\varepsilon}$ detaches from $\kappa =0$ as shown in the inset.
Similarly, a new level $\kappa_{5,\varepsilon}$ detaches from $\kappa =0$ at $a=a_2$. 
 Visually,  horizontal line $\kappa_1^0$ coincides with 
$\kappa_{2,\varepsilon}$
on  interval $0 < a < a_1$\,, with $\kappa_{3,\varepsilon}$
on  interval $a_1 < a < a_2$ and with $\kappa_{4,\varepsilon}$ for $a > a_2$.
 The inset shows how these lines are distinguished.
}
\end{figure}

\begin{figure}[htb]
\begin{center}
\includegraphics[width=0.9\columnwidth]{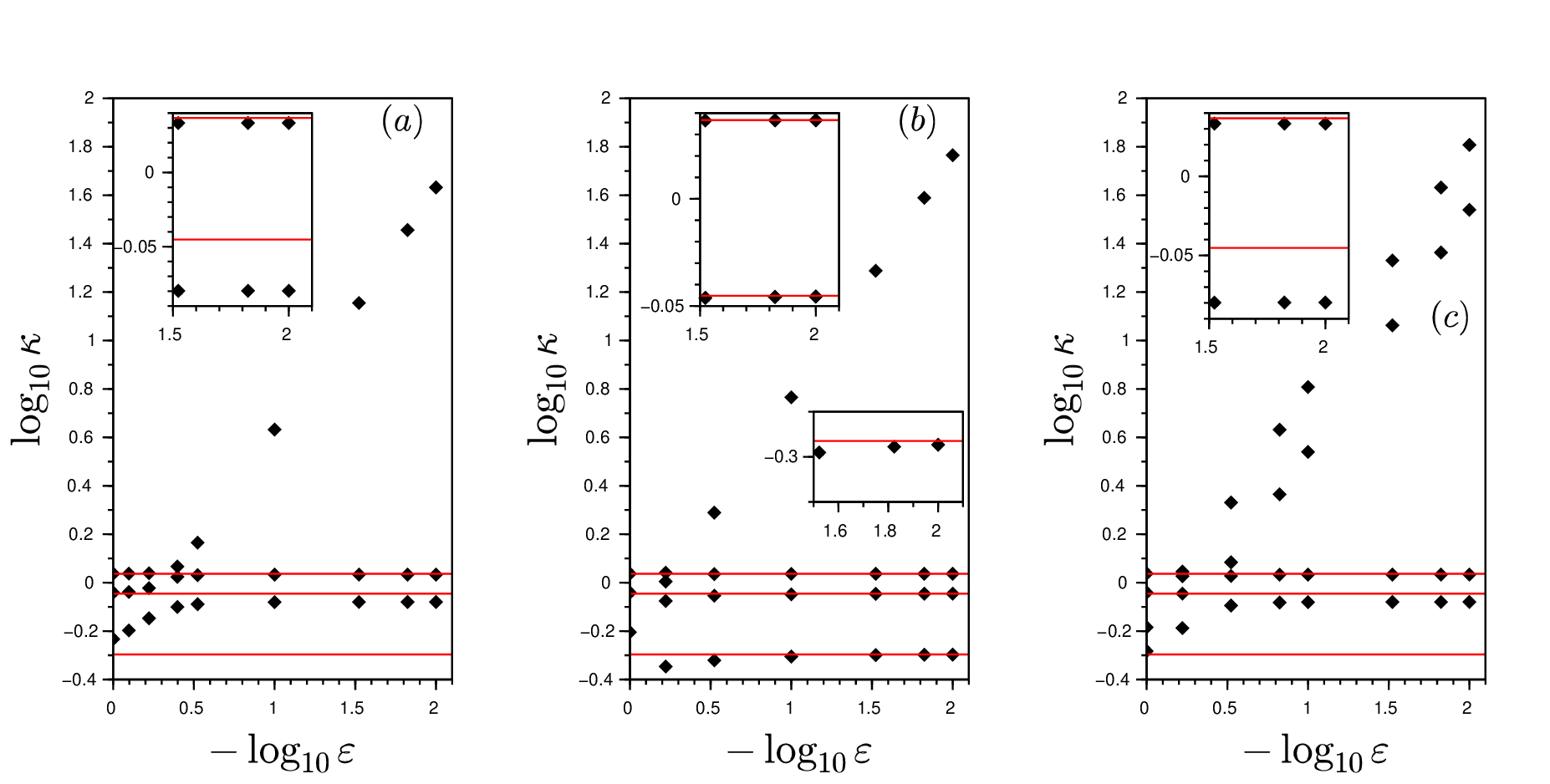}
\end{center}
\caption{\label{fig5} 
The $\varepsilon \to 0$  scenario of the discrete spectrum for the case when the 
$B$-subsystem has the form of the rectangular well that admits 
three bound state levels
$\kappa_1^0$\,, $\kappa_2^0$\,, $\kappa_3^0$ \, ($\kappa_1^0 > \kappa_2^0
> \kappa_3^0$). These levels are shown by red horizontal straight lines. 
The parameter values are $E = 0.01$ eV,  $d =0.2$ eV, $\rho =10$ nm. In panel (a),
at $a =\pi /2\sqrt{d}$ (the middle point of the non-resonance interval 
$0 < a < a_1 = \pi /\sqrt{d}$),  three levels $\kappa_{1,\varepsilon}$,
$\kappa_{2,\varepsilon}$, $\kappa_{3,\varepsilon}$ ($\kappa_{1,\varepsilon} 
> \kappa_{2,\varepsilon} > \kappa_{3,\varepsilon}$) are plotted. Their
$\varepsilon \to 0$ limits are $\kappa_{1,\varepsilon} \to \infty$, 
$\kappa_{2,\varepsilon} \to \kappa_1$, $\kappa_{3,\varepsilon} \to \kappa_2$\,,
where $\kappa_1$ and $\kappa_2$  are solutions to 
Eq.\,(\ref{23}), shown also in the inset. In panel (b), 
at the resonance point $a =a_1=\pi /\sqrt{d}$\,, 
because of detaching a new level from $\kappa =0$, 
the $\varepsilon$-dependent spectrum consists of four levels
($\kappa_{1,\varepsilon} > \kappa_{2,\varepsilon} > \kappa_{3,\varepsilon}
> \kappa_{4,\varepsilon}$). Their $\varepsilon \to 0$ limits are 
 $\kappa_{1,\varepsilon} \to \infty$, 
$\kappa_{2,\varepsilon} \to \kappa_1^0$,  $\kappa_{3,\varepsilon} \to \kappa_2^0$\,,
$\kappa_{4,\varepsilon} \to \kappa_3^0$\,,
where $\kappa_1^0$, $\kappa_2^0$\,, $\kappa_3^0$\,, are solutions to 
Eq.\,(\ref{8}), which are also shown in the inset. 
In panel (c), the $\varepsilon \to 0$ behavior of this four-level spectrum is
plotted for $a =3\pi /2\sqrt{d}$ (the middle point of the non-resonance interval 
$a_1 < a < a_2 = 2\pi /\sqrt{d}$). The $\varepsilon \to 0$ limits of these four 
levels are $\kappa_{1,\varepsilon} \to \infty$, 
$\kappa_{2,\varepsilon} \to \infty$,  $\kappa_{3,\varepsilon} \to \kappa_1$\,,
 $\kappa_{4,\varepsilon} \to \kappa_2$\,,
where again $\kappa_1$ and $\kappa_2$ are the solutions to  Eq.\,(\ref{23}).
}
\end{figure}

In order to examine the bound states of the whole system as functions of 
the prewell parameters and to study the $\varepsilon \to 0$ behavior
if $\tau <1$, we have solved numerically
Eq.\,(\ref{22}) in the special case  when the $B$-potential 
has the form of a rectangular
well with depth $V_b <0$. In this case, we are dealing with a double-well 
potential defined on the interval $0 \le x \le l+\rho+l_b$\,.
A qualitative analysis about the existence of bound states 
 is similar to that given in Ref.\,\cite{zz_jpa}. The results of
 the numerical analysis of the solutions to Eq.\,(\ref{22}), where the elements 
$\lambda_{ij}$'s correspond to a rectangular $B$-well (given by Eq.\,(\ref{9})
with $q= \sqrt{|V_b|-\kappa^2}$ and $l=l_b =y_2- y_1$), 
are illustrated by Figs.\,\ref{fig4} and \ref{fig5}. 
Numerically, the unperturbed spectrum
of the $B$-well, consisting of the solutions to Eq.\,(\ref{8}),
has been  chosen to have three levels of $\kappa$ ($N=3$):  
$  \kappa_1^0 $, $\kappa_2^0$ and $ \kappa_3^0$ ($\kappa_1^0 > \kappa_2^0 > 
\kappa_3^0$). In general, for sufficiently small $\tau < 1$,
 the number of solutions to Eq.\,(\ref{23}) that represent 
 the $\varepsilon \to 0$ limit spectrum of the whole $W$--$B$ structure 
 beyond the resonance set $\Sigma$, is reduced by one becoming $N-1$. 
 Therefore, in our numerical case with $N=3$, the parameter $\tau$ 
 has been chosen in such a way  that  Eq.\,(\ref{23}) 
admits two solutions $\kappa_1$ and $\kappa_2$  ($\kappa_1 > \kappa_2$).
 In more detail, Figs.\,\ref{fig4} and \ref{5} represent 
 the $\kappa$-levels as numerical
solutions to Eq.\,(\ref{22}) at fixed depth $d$ and distance $\rho$. 
At a fixed sufficiently small $\varepsilon$, 
Fig.\,\ref{fig4} illustrates  the dependence of the $\kappa$-solutions 
 on the thickness parameter $a$.  Here, at the
 resonance points $a= a_n : =n\pi/\sqrt{d}$\,,  the new levels are successively
 detaching from the origin $\kappa =0$, so that there are three levels 
on the interval $a_0 < a < a_1$, the additional fourth level 
appears on the interval $a_1 < a < a_2$, and 
the fifth level appears for $a>a_2$\,. In the vicinity of the resonant values $a_n$'s,
the avoided crossings are clearly observed as can be clearly seen in the
inset of Fig.\,\ref{fig4}. Instead of the new levels,
the lowest levels, which depend on $\varepsilon$, tend to infinity as 
$\varepsilon \to 0$.  This limit behavior  is shown in Fig.\,\ref{fig5}, where
the calculations have been carried out for three values of $a$.  
Here, one value is resonant ($a =a_1$) and the other two are  chosen at 
the middles of the non-resonance intervals $0 < a < a_1$ and $a_1 < a < a_2$\,.
The results 
 depicted in both Figs.\,\ref{fig4} and \ref{fig5} can be extended for an 
 arbitrary number  of bound state levels $N$ as follows. Assume that  
 the spectrum of bound states for the $B$-subsystem alone (i.e., at $a=0$) 
 consists of $N$ levels: $ \{ \kappa_1^0\,, \ldots , \kappa_N^0 \}$ ($\kappa_1^0 > 
\ldots > \kappa_N^0$), which are  roots of Eq.\,(\ref{8}).
When a prewell is ``switched on'',  this spectrum on the interval $0 < a < a_1$ is
modified to depend on $\varepsilon$, resulting in the transition 
 $\kappa_1^0 \to \kappa_{1,\varepsilon}\,, \ldots ,\, 
\kappa_N^0 \to \kappa_{N,\varepsilon}$\,. As illustrated by Fig.\,\ref{fig5}(a),
on the interval $0 < a < a_1$\,, we have 
the following limits as $\varepsilon \to 0$: $\kappa_{1,\varepsilon} \to \infty,\,
\kappa_{2,\varepsilon} \to \kappa_1\,,  \ldots , \,
\kappa_{N,\varepsilon} \to \kappa_{N-1}$\,,
so that on this interval the limit spectrum becomes
$ \{ \kappa_1\,, \ldots ,\, \kappa_{N-1} \}$ consisting of the solutions to 
Eq.\,(\ref{23}).

Next, while approaching the next resonance point $a=a_2$\,,  
a new smallest level $\kappa_{N+1,\varepsilon}$  detaches from the origin
$\kappa =0$, as shown in the inset of Fig.\,\ref{fig4}. As a result, at $a = a_2$, 
we have  $N+1$ levels and  the  
$\varepsilon \to 0$ limit is realized as $\kappa_{1,\varepsilon} \to \infty,\,
\kappa_{2,\varepsilon} \to \kappa_1^0\,, \, \ldots , \,
\kappa_{N,\varepsilon} \to \kappa_{N-1}^0\,, \,
\kappa_{N+1,\varepsilon} \to \kappa_{N}^0 $\,, resulting in 
 the spectrum $ \{ \kappa_1^0,\, \ldots , \,\kappa_N^0 \}$. 
 This limiting behavior is illustrated by  Fig.\,\ref{fig5}(b) for the case $N=3$.

Similarly, the transition from the point $a=a_1$ to the interval $a_1 < a < a_2$
leads to the following limit behavior as $\varepsilon \to 0$:  
 $\kappa_{1,\varepsilon} \to \infty,\,
\kappa_{2,\varepsilon} \to \infty, \,\kappa_{3,\varepsilon} \to \kappa_1\,, \,
\ldots , \,\kappa_{N +1,\varepsilon} \to \kappa_{N-1}$\,. For the particular case 
$N=3$, this scenario is demonstrated by Fig.\,\ref{fig5}(c), where  the limiting
discrete spectrum consists of two levels $\kappa_1$ and $\kappa_2$\,. 
Thus, the results described above can be formulated as follows.

{\it Summary 3:  Assume that an unperturbed  $B$-subsystem has a discrete
spectrum $ \{ \kappa_1^0,\,\kappa_2^0,\, \ldots ,\,\kappa_N^0 \}$ composed
of $N$ bound state levels, which are arranged as
$\kappa_1^0 > \kappa_2^0 > \ldots > \kappa_N^0$,
being the solutions to Eq.\,(\ref{8}). Suppose further that $\tau <1$
and the distance $\rho$ provides $N-1$ solutions of Eq.\,(\ref{23}), 
so that the spectrum of these solutions becomes 
$ \{ \kappa_1,\,\kappa_2,\, \ldots , \,\kappa_{N-1} \}$ 
($\kappa_1 > \kappa_2 >\ldots > \kappa_{N-1}$).
Then in the presence of a squeezed rectangular $W$-well, the spectrum of 
the whole $W$--$B$ system preserves on the resonance set $\Sigma$,
consisting of the points $a =a_n = n\pi/\sqrt{d}$, $n \in \N$, exactly 
 the same form as in the case without the well, while beyond
this set, the $W$--$B$ spectrum transforms to $\{ \kappa_1,\,\kappa_2, 
\, \ldots ,\, \kappa_{N-1}\}$ ($\kappa_1 > \kappa_2 >  \ldots > \kappa_{N-1}$).

The $\varepsilon \to 0$ scenario of reconstructing  the $W$--$B$ spectrum 
under the addition of the $W$-well  develops as follows. 
At $a=a_0 =0$, the spectrum is $\{ \kappa_1^0, \ldots , \kappa_N^0\}$.
In the presence of the squeezed $W$-well, this spectrum is modified by adding 
at each $a=a_n$  successively a new smallest 
level that detaches from the origin $\kappa =0$.
Then beyond the  set $\Sigma$, on each interval $a_{n-1} < a < a_n$ 
($n \in \N$), the spectrum of bound states becomes 
$$
\{\kappa_{1,\varepsilon}\,, \,\ldots\, , \,\kappa_{N-1,\varepsilon}\,,\, 
\kappa_{N,\varepsilon}\,,\, \ldots\, , \,\kappa_{n+ N-1,\varepsilon}\}.
$$ 
In the limit as $\varepsilon \to 0$, we have the following convergence
process:
$$
\begin{array}{ccccccccccccccc}
  \kappa_{1,\varepsilon} &  > & \kappa_{2,\varepsilon} & > &\ldots & > &
 \kappa_{n,\varepsilon} & > & \kappa_{n+1,\varepsilon} & > & \ldots & > &
 \kappa_{n+N-1, \varepsilon}   \\
 \downarrow && \downarrow &&&& \downarrow && \downarrow && &&\downarrow  \\
  \infty & & \infty & && & \infty & &\kappa_1 & && &\kappa_{N-1} 
\end{array}
$$
resulting in the limit spectrum 
$ \{ \kappa_1\,,\, \ldots\, ,\, \kappa_{N-1}\} $. 
On the other hand, at each resonant value $a=a_n$\,, the spectrum is 
$$
\{ \kappa_{1,\varepsilon}\,,\, \ldots\, ,\, \kappa_{n,\varepsilon}\,,\,
\kappa_{n+1,\varepsilon}\,,\, \ldots\, ,\, \kappa_{N+n, \varepsilon} \}
$$
and its  $\varepsilon \to 0$ limit reads
$$
\begin{array}{ccccccccccccccc}
  \kappa_{1,\varepsilon} &  > & \kappa_{2,\varepsilon} & > &\ldots & > &
 \kappa_{n,\varepsilon} & > & \kappa_{n+1,\varepsilon} & > & \ldots & > &
 \kappa_{n+N, \varepsilon}   \\
 \downarrow && \downarrow &&&& \downarrow && \downarrow && &&\downarrow  \\
  \infty & & \infty & && & \infty & &\kappa_1^0 & && &\kappa_{N}^0 
\end{array}
$$
 leading finally to the limit spectrum
$ \{ \kappa_1^0\,,\, \ldots\, ,\, \kappa_{N}^0\} $. 
}
 
 \section{Concluding remarks}
 
 Thus, we have investigated the influence of an extremely thin potential
  well on the spectrum of the background layer with the potential profile
 of an arbitrary shape. For these studies to be implemented, 
  it was necessary to determine correctly the zero-range limit
 of a thin well in the sense of point interactions. More precisely, 
while the typical $\varepsilon \to 0$ limit for the barrier potential 
$V(x)$ in Eq.\,(\ref{1})
is $\varepsilon^{-\nu}V(x/\varepsilon)$ with $\nu =1$, in the case of a well,
it turns out to be at $\nu =2$. This is a key point of our approach
for the study of the role of a squeezed potential well in the transmission 
of quantum particles through a barrier. The other interesting question is
how the presence of such a well disturbs the spectrum of bound states 
in any background system located at some finite (sufficiently close) distance. 

In order to implement the $\varepsilon \to 0$ limit explicitly, we have chosen 
the potential well $V(x)$ in a rectangle-like form. In the squeezing limit as 
$\varepsilon \to 0$, both the cases $\nu =1$ and 
$\nu =2$ demonstrate in Fig.\,\ref{fig1} the crucial difference of the transmission 
across this well, from which it immediately follows that  the approximation with
$\nu =2$ turns out to be much more appropriate than the typical $\delta$-like
representation with $\nu=1$. Note that the potential  
$\varepsilon^{-2}V(x/\varepsilon)$ has no limit as $\varepsilon \to 0$. However,
the resulting point interaction is well-defined because it satisfies the
two-sided (at $x=\pm \,0$) boundary conditions on the wave function $\psi(x)$ 
and its derivative $\psi^\prime(x)$ \cite{adk}.
 Indeed, on the resonance set $\Sigma$, the corresponding connection matrix reads
  $(-1)^n I$, where $I$ is the identity matrix and $n=1,\,2,\, \ldots \,$. 
  Beyond the set $\Sigma$, the boundary conditions are of the Dirichlet type:   
  $\psi(\pm \,0)=0$.

The whole system studied in the present paper consists
of two separated (by some distance) parallel homogeneous layers, 
where an extremely thin potential well  (referred to as a $W$-subsystem or a prewell) 
is auxiliary to a background regular structure (called a $B$-subsystem or a $B$-layer), 
the potential profile of which has an arbitrary shape. It is remarkable, as illustrated by 
Fig.\,\ref{fig3},  that the decrease in prewell thickness  
makes the resonant transmission through the $B$-layer more pointed
and significantly reduced between the resonance points. Therefore, adjusting the size
parameters of the prewell, it is possible to control the net current 
in the whole system. This effect can be realized as a {\it filter} in nanodevices. 
 
 An interesting scenario as $\varepsilon \to 0$  has been observed 
 when a squeezing prewell  is incorporated nearby the regular $B$-layer having a non-empty 
  spectrum of bound states.  At each  point $a_n = n\pi/\sqrt{d}$, 
  $n=1,\,2,\, \ldots\,,$ from the resonance set 
  $\Sigma$ that corresponds to the squeezed prewell, a new 
  bound state level in the whole $W$--$B$ system is detaching from 
the origin $\kappa =0$ at finite values of $\varepsilon >0$.
However, in the limit as $\varepsilon \to 0$, equally the same number
of the lowest levels escape to infinity. As a result, at resonant points
of the set $\Sigma$, the prewell does not affect the spectrum of the $B$-subsystem.
Outside $\Sigma$, the spectrum is modified obeying Eq.\,(\ref{22}) and being the same
along all the intervals $a_{n-1}< a <a_n$\,, $n=1, \,2,\, \ldots\,.$ 
 
\section*{CRediT authorship contribution statement}  
{\bf Yaroslav Zolotaryuk:} Conceptualization, Formal analysis, Investigation, Software, Validation, Visualization, Writing – review \& editing.
 {\bf Alexander V. Zolotaryuk:} Conceptualization, Formal analysis, Investigation, Methodology, Writing – original draft, Writing – review \& editing.
 
\section*{Declaration of competing interest} 
 The authors declare that they have no known competing financial interests or personal relationships that could have appeared to influence the work reported in this paper.
 
\section*{Data availability} 
 
 Data will be made available on request.
 
 \section*{Acknowledgements}
 
 We would like to thank the Armed Forces of Ukraine
for providing security to perform this work. The authors 
acknowledge ﬁnancial support from 
the National Academy of Sciences of Ukraine by projects No.~0122U000887 (Y.Z.) and
No.~0122U000888 (A.V.Z.). 
This work was also supported by the Simons Foundation.

\nocite{*}
\bibliographystyle{unsrt}

\end{document}